\documentclass[prc,article,twocolumn,amssymb]{revtex4}

\usepackage{graphics}
\usepackage{epsfig}

\begin{document}

\title{Realistic shell-model calculations and exotic nuclei}

\author{A. Gargano$^a$, L. Coraggio$^a$, A. Covello$^{a,b}$, N. Itaco$^{a,b}$}

\address{$^a$Istituto Nazionale di Fisica Nucleare,\\
Complesso Universitario di Monte S. Angelo, I-80126 Napoli\\
$^b$Dipartimento di Fisica, Universit\`{a} di Napoli Federico II,\\
Complesso Universitario di Monte S. Angelo, I-80126 Napoli}

\date{\today}

\begin{abstract}
This paper  presents a short overview  of the shell-model approach with realistic effective interactions to the study of exotic nuclei.
We first  give a sketch of the current state of the art of the theoretical framework of this approach,  focusing on the main ingredients and most relevant recent advances. Then, we present some selected results for neutron-rich nuclei in various mass regions, namely oxygen  isotopes, $N=40$ isotones,   and nuclei around $^{132}$Sn,  to show the merit  as well as the limits of these calculations.
\end{abstract}

 \maketitle

\section{Introduction}

The shell model has long proved to be a main key  to the  understanding  of nuclear structure. It provides the theoretical framework for a microscopic description of nuclear properties which is essentially based on the use of effective interactions. In fact, as  is well known, within the shell-model approach only the particles outside a core made up of filled shells  (valence particles) are  considered to be active, and calculations are performed in a truncated Hilbert space, the so-called model space. The shell-model Hamiltonian acting only between the valence particles should, therefore, account for the neglected degrees of freedom, namely those of the core particles as well as of the excitations of valence particles above the chosen model space.

To this end, one can resort to empirical interactions,  {\it i.e.}  interactions containing adjustable parameters or obtained by treating the matrix elements themselves as free parameters. In  both cases  fitting procedures  to reproduce the experimental  data are required.  Empirical interactions have been used in a  number of shell-model calculations, providing in most cases  a successful description of a  variety of nuclear phenomena.

Clearly, a more fundamental approach to the shell model consists in starting  from the interaction between free nucleons and constructing the Hamiltonian  by means of many-body techniques, which leads  to what is called  ``realistic effective interaction''.  This alternative way, based on a microscopic derivation of the shell-model effective interaction, has the great advantage  that no adjustable parameter is needed   and establishes a  bridge between effective  shell-model interactions and   underlying nuclear forces. Significant progresses have been made along this line in the last two decades, and realistic shell-model calculations  have been shown to provide an accurate description of nuclear structure properties for nuclei in various mass region both   close to and far  from the stability  valley.

In this context, it should be mentioned that in many shell-model calculations, in particular those aiming at interpreting  the new experimental data  obtained at Radioactive Ion Beam (RIB) facilities,    use has also been  made  of semi-empirical interactions, consisting in modified versions of realistic effective interactions. Here, however, we focus only on shell-model calculations employing genuine realistic effective interactions  and  give a brief description  of the current status of this approach. We firstly review the main steps and the relevant recent developments  involved in  the perturbative technique used to  derive the shell-model interaction from the bare nuclear potential. Then, we report some results we have obtained  for various neutron-rich nuclei, as oxygen isotopes \cite{Coraggio07,Coraggio11}, $N=40$ isotones \cite{Coraggio14},  and nuclei around $^{132}$Sn \cite{Coraggio13}, to show  the practical value of this approach as well as its limits. 

The basic ingredient of realistic shell-model calculations is the bare nuclear potential, for which, as is well known, there are various reliable models.  However, we shall not discuss this point here. In the calculations concerned with  the present contribution we have used two modern nucleon-nucleon ($NN$) potentials, the CD-Bonn \cite{Machleidt01} and N$^{3}$LOW potential \cite{Coraggio07}, which fit equally well the $NN$ scattering data.  In particular, the calculations for $N=40$ isotones   and nuclei around $^{132}$Sn 
 have been  performed with the former while results for oxygen isotopes have been obtained with the latter. It is worth recalling that, owing to its   strong short-range repulsive behavior, the CD-Bonn potential cannot be used directly  in deriving the effective interaction within the framework of a perturbative approach. In other words, it must  be first renormalized, which is done by constructing  a low-momentum potential  $V_{\rm {low-k}}$ defined within a cutoff momentum $\Lambda$~\cite{Bogner02}. This is a smooth potential  which preserves
exactly the onshell properties of the original one. As concerns  the N$^{3}$LOW potential, this is derived from the chiral perturbation theory  with a sharp momentum cutoff at 2.1 fm$^{-1}$ . It is conceived therefore as a low-momentum potential and no renormalization procedure is needed.

Finally, we would like to point out that our effective interactions are all based only on bare $NN$ potentials, namely  three-body forces are not explicitly considered. The study of  the role of  three body forces in nuclear structure  has recently attracted great theoretical interest. Their effects have  been  evidenced  within the framework of {\it ab initio} approaches for few nucleon systems. Until now, however, no shell-model calculation has been performed with an effective Hamiltonian derived by treating on equal footing both $NN$ and 3$N$ forces. This, which  would imply the appearance of  further core-polarization effects on the  one- and two-body components as well as of an effective three-body term, is quite a complex project.   Actually, only  first-order contributions of the normal-ordered one- and two-body parts of 3$N$ forces have been taken explicitly into account.  In connection with the role of 3N forces, however, it should be mentioned that in a recent paper \cite{Ekstrom13}  an optimized $NN$ interaction derived from the chiral effective field theory has been constructed, which seems to account for many aspects of nuclear structure without explicitly including 3$N$ forces. 

\section{Theoretical framework}

Let us start with the Schr\"odinger equation for a system of A nucleons interacting via two-body 
forces

\begin{equation}
H\Psi_{\alpha}= E_{\alpha} \Psi_{\alpha}, \label{Schr} 
\end{equation}

\noindent
 where

\begin{equation}
H = T+V_{\rm NN},
\label{defh}
\end{equation}

\noindent
$T$ being the kinetic energy and $V_{\rm NN}$ a low-momentum  two-body potential,
obtained through the $V_{\rm low-k}$ procedure \cite{Bogner02} or purposely constructed to have a smooth perturbative behavior, with the addition of the Coulomb force for protons.

Now, by introducing an auxiliary one-body potential $U$  
the Hamiltonian~(\ref{defh}) can be  written as

\begin{equation}
H= (T+U)+( V_{NN}-U)= H_{0}+H_{1},
\label{eq1}
\end{equation}

\noindent
namely as a one-body component $H_0$, which describes the independent motion of 
the nucleons, and a residual interaction $H_1$. 

The effective Hamiltonian, $H_{\rm eff}$, is defined through 
the model-space eigenvalue problem
 
\begin{equation}
H_{\rm eff}P| \Psi_\alpha\rangle = H_{0}P| \Psi_\alpha\rangle + H_{1}^{\rm eff}P
 | \Psi_\alpha\rangle=E_\alpha P \Psi_\alpha, \label{defheff} 
\end{equation}

\noindent
where  the  $E_\alpha$  and the corresponding $\Psi_\alpha$ are a subset of the
eigenvalues  and eigenfunctions of the original Hamiltonian. Clearly, 
$H_{\rm eff}$  acts only on the model space defined in terms of the eigenvalues of  $H_{0}$ through the projection operator $P$.

A well-established approach to the determination  of the  effective Hamiltonian is given by the 
$\hat Q$-box folded-diagram expansion. A detailed description of this approach can be found in Refs.~ \cite{Coraggio09,Coraggio12},  so we will not touch upon it here. We would 
like, however,  to highlight the main points involved in the derivation of 
$H_{\rm eff}$,  so as to make clear the present stage of development.

Within a degenerate model space, $PH_{0}P= \epsilon_{0}$, iterative techniques~\cite{Suzuki89} can be used to construct the effective Hamiltonian. These, as the Krenciglowa-Kuo (KK) and the Lee-Suzuki (LS) ones,  are based on an expansion of $H_{1}^{\rm eff}$ in terms of the  $\hat Q$-box and its derivatives, the  $\hat Q$-box being defined as
\begin{equation}
\hat Q(\epsilon)= P H_{1} P + PH_{1} Q \frac{1}{\epsilon-QHQ} QH_{1}P, \label{defq}
\end{equation}

\noindent
where the  operator $Q$ is the complement of $P$.

Once the  $\hat Q$-box is calculated,  we derive $H_{1}^{\rm eff}$ by means of the LS technique, which yields converged results  after a small number of iterations.
The calculation of the $\hat Q$-box, however, is  the  most critical step of our procedure. This is performed by writing the term  $1 /({\epsilon- QHQ}$) in equation~(\ref{defq}) as a power series,  which leads to a  perturbative calculation to be performed under some approximations. 
A diagrammatic representation of the $\hat Q$-box, including one- and two-body diagrams up to third order in the interaction, is given in Ref. \cite{Coraggio12}.  Clearly, only diagrams up to a finite order can be included and the state-of-the-art calculations do not go beyond the third order.   One should also consider that the evaluation of the diagrams composing the $\hat Q$-box requires, in principle, a summation over all the states of the $Q$ space. The truncation of this infinite space is, therefore, another  source of approximation. Both the order-by-order  and the intermediate-state convergence of the effective interaction expansion are briefly reviewed in \cite{Coraggio09,Covello10} where references to previous works  are given, while in \cite{Coraggio12} they are discussed in detail focusing on $p$-shell
 nuclei. In section 3,  results of shell-model calculations with effective interactions derived by including diagrams up to second as well as third order are reported and compared with experimental data.

In concluding this outline of the theoretical framework, it is worth mentioning two other points entering our procedure.
Our effective Hamiltonian is  derived for a two-valence-particle nucleus but is then used for systems  with a larger number of valence particles. This means that we only include one- and two-body forces neglecting higher-body terms, which  arise as an effect of the nuclear medium even if  an $NN$ potential is used.  The one-body force gives the theoretical single-particle energies as resulting  from the sum of the eigenvalues of $H_0$ and the one-body contributions of $H_{1}^{\rm eff}$. In most realistic shell-model calculations, however, these energies are replaced by values taken from  experiment. Furthermore, we note that the
perturbation expansion is performed for $H_{1}=V_{\rm NN} - U$,  which gives rise to 
 $(V-U)$-insertion diagrams (see \cite{Coraggio12}), which are in general neglected with the exception of the first-order ones.  As a matter of fact, these are exactly zero only when taking for $U$ a self-consistent Hartree-Fock  potential, whereas an harmonic oscillator potential is generally used.
In the next section we shall present results obtained by using both theoretical and experimental single-particle energies as well as by including or not  $(V-U)$-insertion diagrams beyond the first order.

\section{Results for various mass regions}
We discuss and compare with experiment results for O isotopes, $N=40$ isotones, and $^{132}$Sn neighboring nuclei, which have been obtained within the shell-model framework using realistic effective interactions.  Most of the results presented in this section have already been given in previous works \cite{Coraggio11,Coraggio14,Coraggio13},  but our aim here is  to give a panoramic view of them in order to  show the ability of these effective interactions to describe neutron-rich nuclei in different mass regions.  

\subsection{Oxygen isotopes}

One of the challenging problems in modern nuclear structure studies concerns the location of the neutron drip line, namely the  limit of existence of neutron-rich systems. In this context, the oxygen isotopes play an important role. The drip line for $Z=8$ is quite close to the stability valley, in contrast with the situation that occurs for other isotopes in the same mass region. As a matter of fact, the limit for oxygen isotopes  is established at $N=16$, the last stable one being $^{18}$O. Several calculations \cite{Otsuka10,Hagen12,Hergert13} have been recently performed  suggesting  that the explanation for this limit resides  in $3N$ forces.

The effective Hamiltonian for oxygen isotopes is derived for the $sd$ space with $^{16}$O as inert core starting from the $N^3$LOW potential with a sharp cutoff at 2.1 fm$^{-1}$, which is a low-momentum realistic $NN$ interaction derived from the chiral perturbation theory. All diagrams up to third order are taken into account  in the calculation of the $\hat Q$-box including the  $(V-U)$-insertion diagrams. As regards the energies of the three single-particle orbitals of the $sd$ space, we employ the theoretical  values, as arising from the one-body contributions of $H_{\rm eff}$.

\begin{figure}[h]
\begin{minipage}{18.5pc}
\vspace{-1.5pc}
\includegraphics [width=20pc] {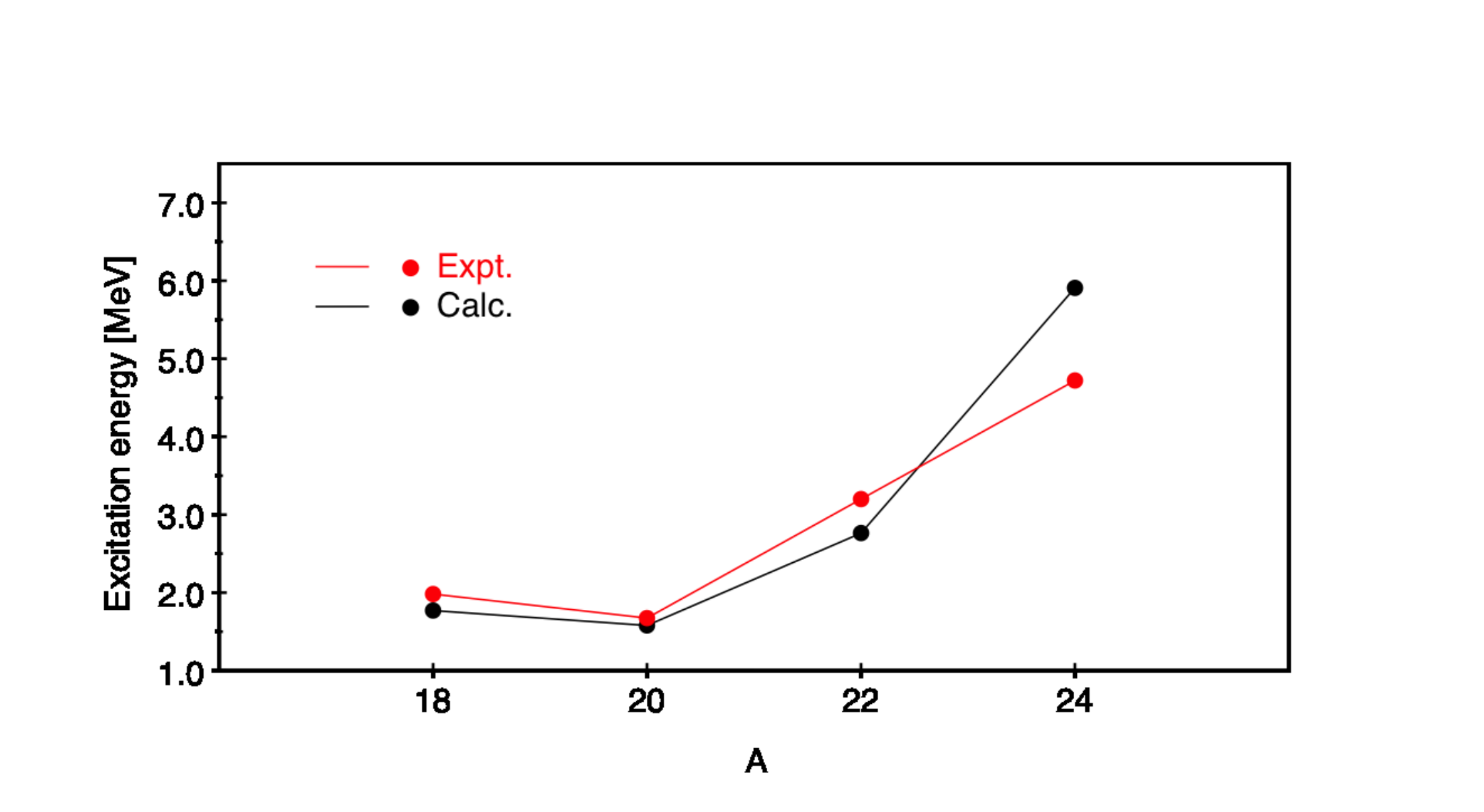}
\caption{\label{fig1_AG} (Color online) Excitation energies of yrast $2^+$ states in O isotopes from $A=18$ to 24 .}
\end{minipage}\hspace{1.pc}
\begin{minipage}{18.5pc}
%\vspace{-0.5pc}
\includegraphics [width=20pc] {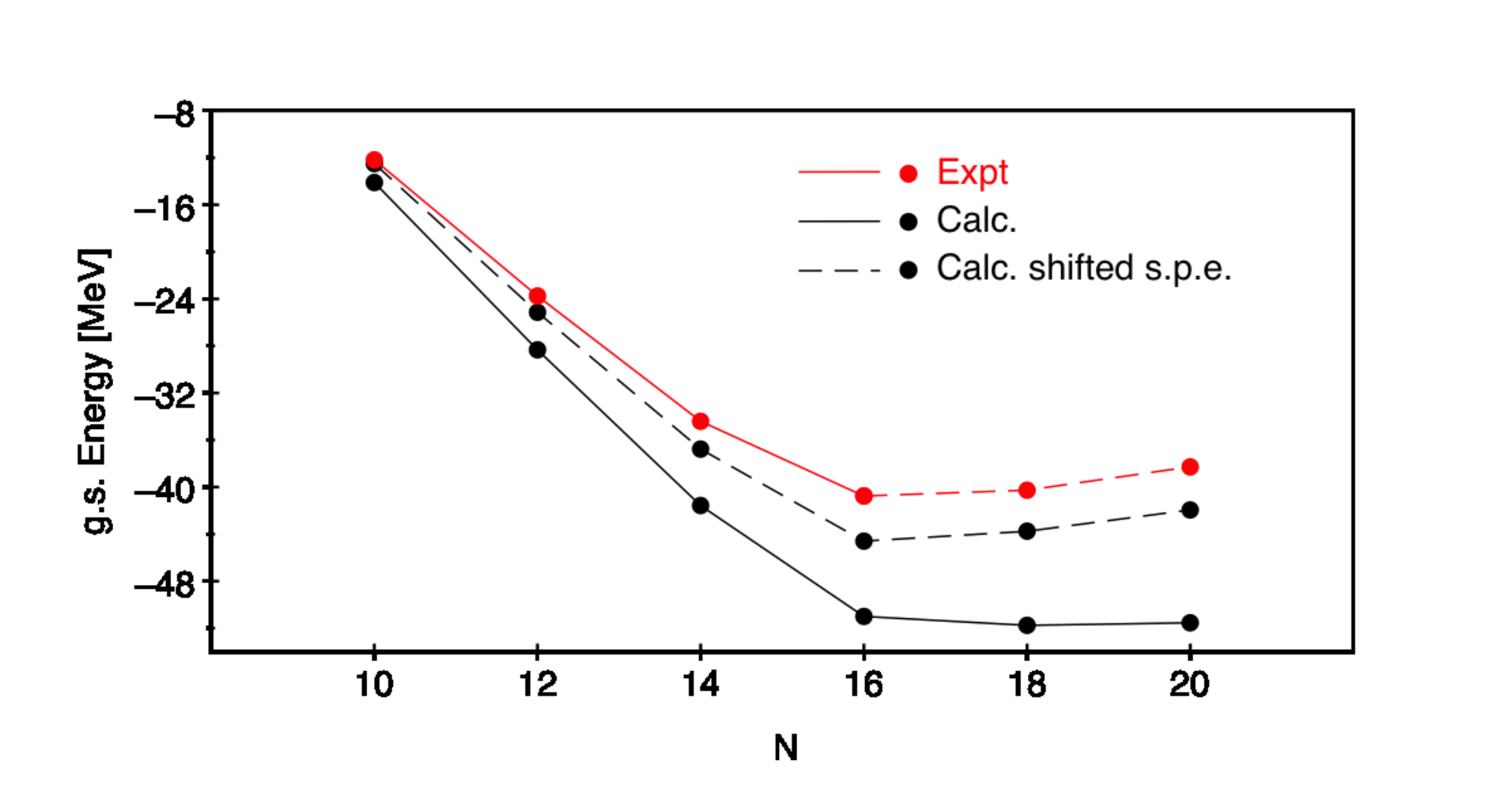}
\caption{\label{fig2_AG} (Color online) Ground-state energies for O isotopes from $N=10$ to 20 (see text for details).}
\end{minipage} 
\end{figure}

In figure~\ref{fig1_AG}, the calculated excitation energies of the yrast $2^{+}$ states are compared with the experimental values~\cite{NNDC}. We see that our predictions are in very good agreement with the observed energies.  In particular, we reproduce the  rise from $A=20$ to 22 as well as that from $A=22$ to 24 which are related to the $N=14$ and 16 subshell closure, respectively.
 
To test the ability of our interaction to reproduce the location of the drip line for oxygen isotopes, we have calculated the ground-state  (g.s.) energies, which are compared in
figure~\ref{fig2_AG} with the experimental values~\cite{AMU}. The theoretical curve lies below the experimental one, with a discrepancy increasing with the number of neutrons. Actually, our calculations overestimate the experimental g.s. energies and fail to predict $^{26}$O and $^{28}$O as unbound nuclei. In the same figure, however, we show the results that are obtained by an  upshift (427 keV) of the calculated single-particle  spectrum so as to reproduce the experimental g.s. energy of $^{17}$O relative to $^{16}$O. The new curve moves up coming closer to  the experimental one  and shows the right slope from $N=16$ on. This highlight some inaccuracy in our one-body effective Hamiltonian which may be  traced to the lack of three-body forces. 

\subsection{$N=40$ isotones}
We now present and discuss our results for the $N=40$ isotones Ca, Ti, Cr, Fe, and Ni. Before doing so, however, a few comments are in order. The experimental behavior of the excitation energy of the $2^+$ yrast state in Ni isotopes, and in particular the sizable increase in $^{68}$Ni with respect to the two neighboring even isotopes, evidences a subshell closure at N=40. For a decreasing number of protons this closure disappears and the onset of a collective behavior is observed.  This issue is currently the subject of great experimental and theoretical interest (see~\cite{Lenzi10,Rother11,Baugher12,Crawford13} and references therein), and the appearance of the collective behavior has been traced  to the correlations between
the quadrupole-partner neutron orbitals $0g_{9/2}$ and $1d_{5/2}$~\cite{Lenzi10}.

\begin{figure}[h]
\begin{center}
\includegraphics[width=20pc]{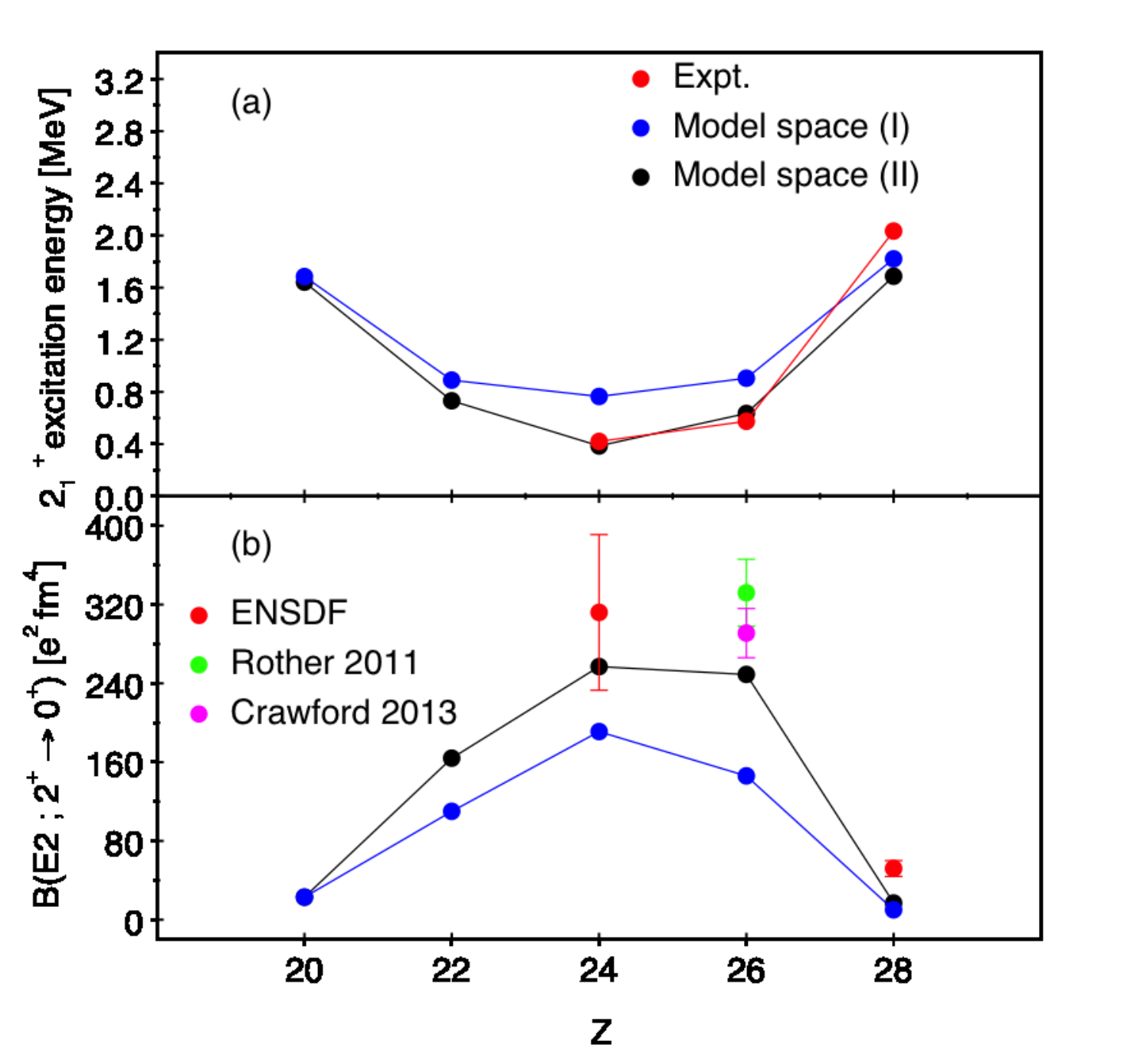}
\end{center}
\caption{\label{fig3_AG} (Color online) (a) Experimental (ENSDF \cite{NNDC}, Rother \cite{Rother11}, Crawford \cite{Crawford13}) and calculated excitation energies of the yrast $2^+$ states and (b) $B(E2; 2^{+}_{1} \rightarrow 0^{+}_{1})$ for the $N=40$ isotones.}
\end{figure}

With the aim  of directly testing the role played by the neutron $1d_{5/2}$ orbital, we have  performed realistic shell-model calculations taking $^{48}$Ca as inert core and considering two different model spaces. The first one, model space (I), is spanned by the proton $0f_{7/2}$ and $1p_{3/2}$ orbitals and by the neutron $1p_{3/2}$, $1p_{1/2}$, $0f_{5/2}$, $0g_{9/2}$ orbitals while the second one, model space (II), includes the same orbitals with the 
addition of the neutron  $1d_{5/2}$ one. We calculate the effective interaction starting from the CD-Bonn $NN$ potential renormalized by way of the $V_{\rm low-k}$ approach with a cutoff momentum $\Lambda=2.6$~fm$^{-1}$. As for the O isotopes, all diagrams up to  third order are taken into account  in the calculation of the $\hat Q$-box, but at variance with the previous case the single-particle energies are determined using experimental data. More details on our effective Hamiltonian, including the values of the single-particle energies and a list of the two-body matrix elements in the model spaces (I) and (II) are given in Ref.~\cite{Coraggio14}.

In figure~\ref{fig3_AG}, we report the excitation energies of the yrast $2^+$  states  and the $B(E2; 2^{+}_{1} \rightarrow 0^{+}_{1})$ transition rates for the  $N=40$ isotones as a function of $Z$.  As regards the energies, we see that both calculated curves reproduce well the observed behavior, but the inclusion of the neutron $1d_{5/2}$ orbital is essential for a quantitative agreement with the experimental data. The  role of this orbital appears  even more relevant when looking at the $B(E2)$ transition rates. Only the $B(E2)$'s for $^{64}$Cr, $^{66}$Fe, and $^{68}$Ni are experimentally known, and these values for the former two nuclei are significantly underestimated by our calculations with model space (I), while this  is not the case when model space (II) is used. Our results, therefore, confirm the connection between the $1d_{5/2}$ orbital and the appearance of collectivity in Ti, Cr, and Fe, as evidenced by the lowering of the $2^{+}_{1}$ state and the increase in  the corresponding $B(E2)$. To better understand this point,  we show in  table~\ref{tab1} the occupation numbers of  the proton $\pi f_{7/2}$ and neutron $\nu g_{9/2}$ and $\nu d_{5/2}$ orbitals for the  ground state of Ti, Cr, Fe, and Ni.

\begin{table}[h]
\caption{\label{tab1} Occupation numbers of the proton $\pi f_{7/2}$ and neutron $\nu g_{9/2}$ and $\nu d_{5/2}$ orbitals for the ground state of the $N=40$ isotones (see text for details).} 
\begin{ruledtabular}
\vspace{0.6cm}
\begin{tabular}{llcccc}

& & Ti & Cr & Fe & Ni \\

\colrule
$\pi f_{7/2}$ & (I)  & 1.87 &  3.64 & 5.61 & 7.89 \\
                     & (II) & 1.69 &  3.31 & 5.31 & 7.85 \\
 $\nu g_{9/2}$& (I)  & 3.23 &  2.88 & 1.75 & 0.42 \\
                      & (II)  & 3.72 &  3.73 & 2.81 & 0.47 \\
$\nu d_{5/2}$ & (II) & 0.36 &  0.57 & 0.33 & 0.05 \\

\end{tabular}
\end{ruledtabular}
\end{table}

\begin{figure} [h]
\begin{center}
\includegraphics[width=20pc]{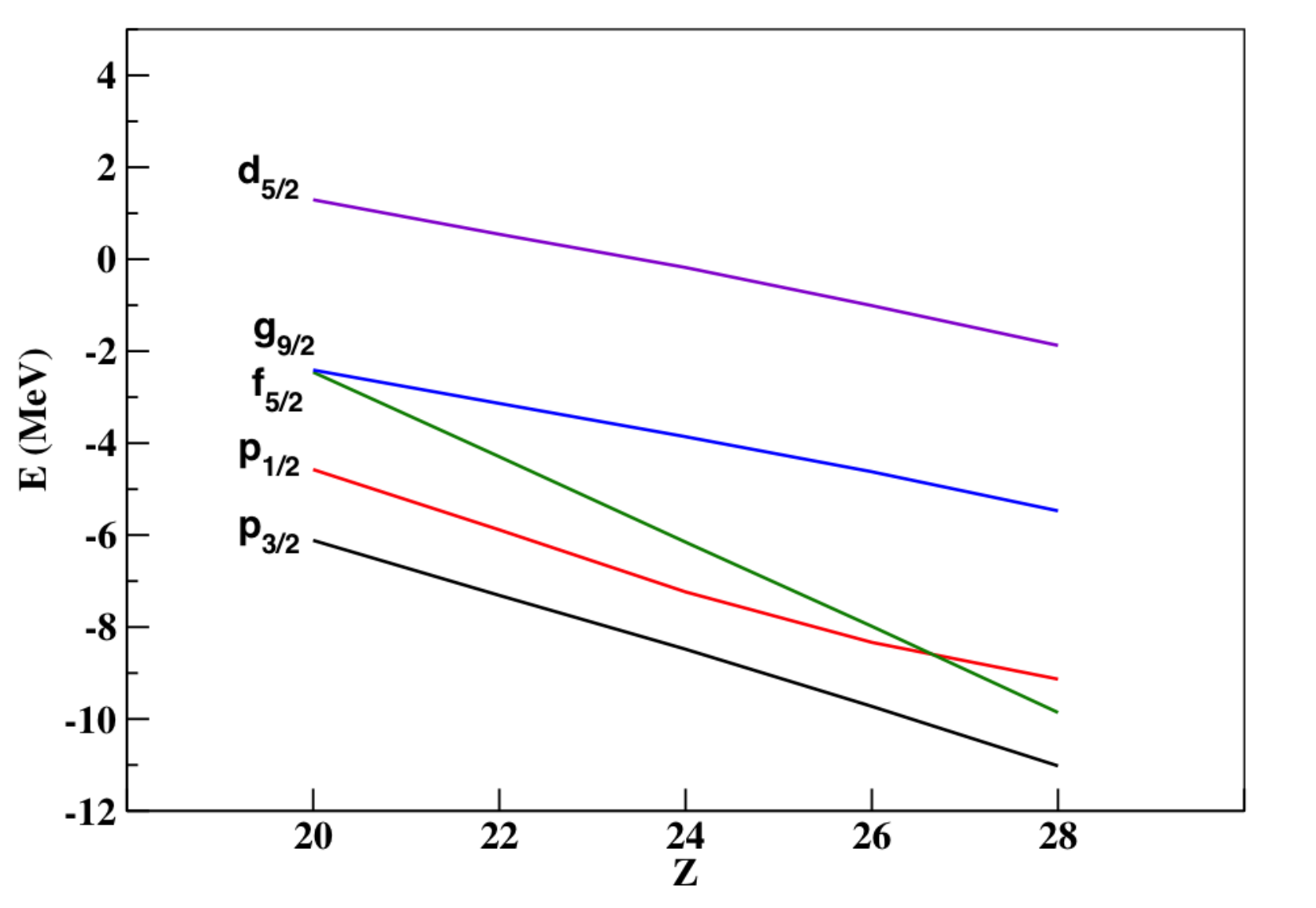}
\end{center}
\caption{\label{fig4_AG} (Color online) Calculated effective  single-neutron energies for the $N=40$ isotones.}
\end{figure}

We see that there are no substantial differences between the occupation numbers  obtained with model spaces (I) and (II) for Ni. In both cases, a very low occupancy of protons (0.11-0.15) is found  above the $0f_{7/2}$ orbital as well as of neutrons (0.42-0.52) above the $0f_{5/2}$ orbital, the latter result implying a clear manifestation of  subshell closure at $N=40$. Note that, with model space (II),  only 0.05
neutrons occupy the $1d_{5/2}$ orbital which  gives reason for its minor role in the description of $^{68}$Ni. 
However, when decreasing the number of protons the two model spaces lead to significantly different results. In fact, when going from model space (I) to model space (II) we find an increase in the occupation number of the neutron $0g_{9/2}$ orbital as well as a depletion of the proton $0f_{7/2}$ orbital. From table~\ref{tab1}, we also see an increase in the occupation number of the  neutron $1d_{5/2}$ orbital for Ti, Cr, and Fe with respect to Ni. In other words, neutron excitations above the $fp$ orbitals are favored for protons numbers below the 28  closed shell. This  may be traced to a reduction of the neutron $0g_{9/2}-0f_{5/2}$ gap  as well as to  the quadrupole-quadrupole component of the effective interaction acting between the $0g_{9/2}$ and $1d_{5/2}$ orbitals. The former argument is clearly evidenced in figure~\ref{fig4_AG}, where we report  the behavior of the effective single-neutron energies as a function of $Z$.

\vspace{0.5cm}
\subsection{$^{132}$Sn neighboring}
In the last ten years or so, nuclei in the mass region around $^{132}$Sn have become accessible to experimental studies thanks to  new RIB facilities  and the  development  of sophisticate detection
techniques. A  large amount of experimental information has been acquired, but data for nuclei with $Z \sim 50$ and $N > 82$ still remain scarce. A further step in this direction will be  certainly made  with the next generation of RIB facilities. The available data for $N>82$ nuclei have not evidenced changes in the shell structure, as it is was instead the case for the lighter nuclei discussed above. However, some  anomalies have been observed. A notable one, which have posed interesting questions is, for instance, the asymmetric behavior with respect to $N=82$ of the excitation energy of the yrast $2^+$ state in both Sn and Te isotopes.

We have conducted several studies~\cite{Coraggio9b,Covello11,Danchev11,Coraggio13b} on nuclei of this region, in all of them taking $^{132}$Sn as  closed core and assuming that proton particles and neutron holes occupy the five orbitals of the 50-82 shell, while neutron particles are in the six orbitals of the 82-126 shell.  The single-particle and single-hole energies have  been  determined from  experiment and the two-body effective interaction has been  derived from the CD-Bonn $NN$ potential renormalized by means of the $V_{\rm low-k}$ potential with $\Lambda=2.2$~fm$^{-1}$. As for the calculation of the $\hat Q$-box, all diagrams up to second order have been taken into account, except the $(V-U)$-insertion diagrams which are limited to  first order.  All these studies,  focused on energy spectra and electromagnetic properties, and more recently on spectroscopic factors and atomic masses, have led to results in very good agreement with experiment. 

A main outcome  of our work is that the pairing force plays a key role  in determining  the various properties of $^{132}$Sn neighbors, as  for instance the decrease in energy of the $2^{+}$ state in $^{134}$Sn and $^{136}$Te. Our effective interaction, in fact, generates a pairing force between two neutrons in the 82-126 shell which is significantly weaker than that between   two neutron holes or two proton particles in the 50-82 shell.
Let us take, for instance, the  $(\nu f_{7/2})^{2}$ configuration. The corresponding $J=0^+$ matrix element is about -0.6 MeV to be compared to the value of -1~MeV or less for the 
$(\nu h_{11/2})^{-2}$ and $(\pi g_{7/2})^{2}$ configurations. This issue is discussed in \cite{Covello13}, where we have investigated the origin of the pairing force within our microscopic framework  and found that the above differences result from  a large reduction of the core-polarization contributions to the neutron effective interaction for $N>82$. 
It should be mentioned that the role of  the pairing force in this mass region was also recognized in Refs.\cite{Terasaki02,Shimizu04}.

\begin{figure} [h]
\hspace{1.5cm}
%\begin{center}
\includegraphics[width=23pc]{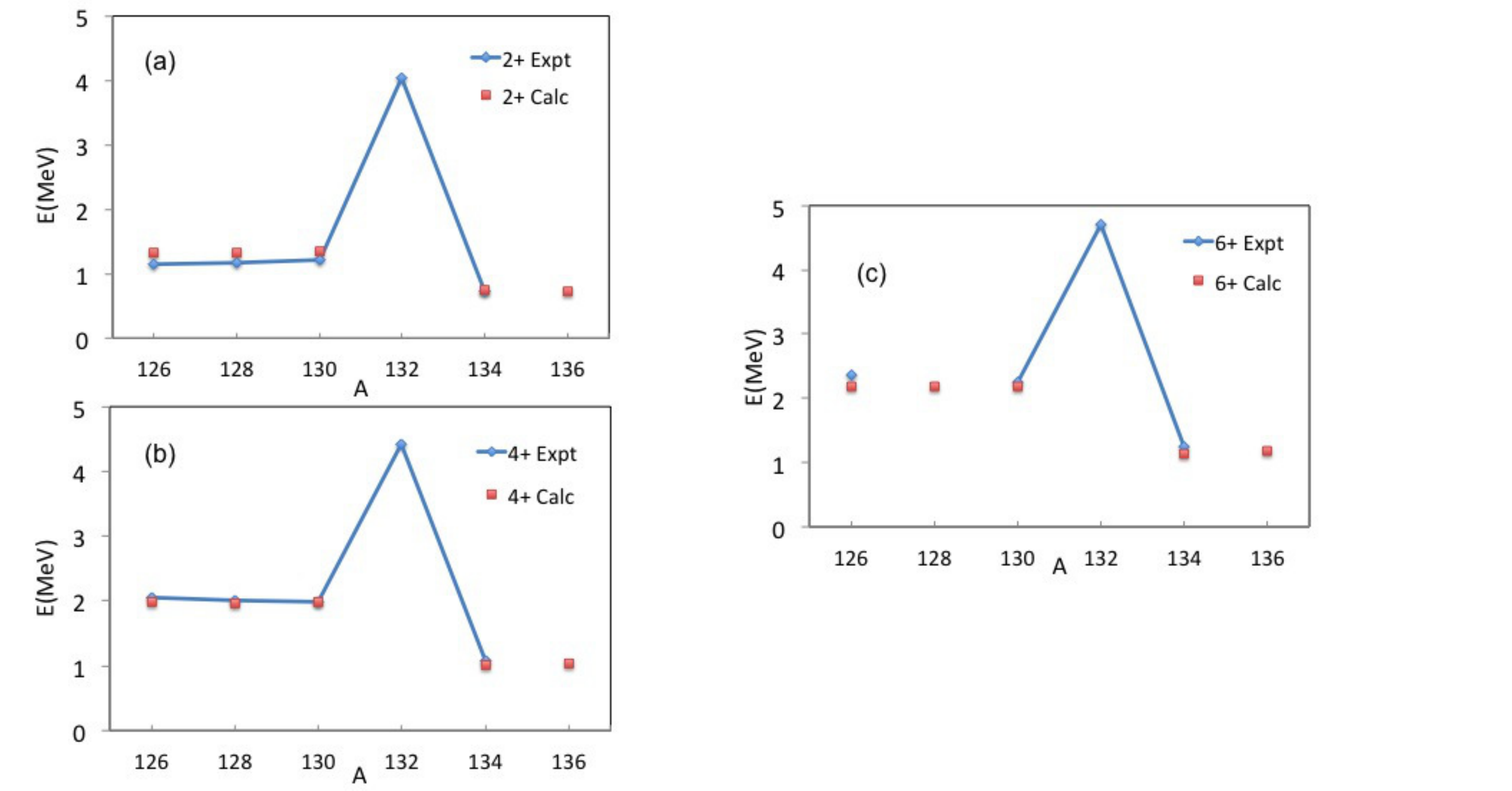}
%\end{center}
\caption{\label{fig5_AG} (Color online) Calculated and experimental~\cite{NNDC} excitation energies of the yrast (a) $2^+$, (b) $4^+$, and  (c) $6^+$ states in tin isotopes from $A=126$ to 136.}
\end{figure}

To illustrate the quality of our results, we focus on the low-energy spectra and $B(E2)$ transition rates in  Sn and Te isotopes. Then, as a final example, we discuss the odd-even staggering (OES) of binding energies for $N= 81$ and 83 isotones to show the predictive power 
 of our approach in the calculation of binding energies.

\begin{figure} [h]
\begin{center}
\includegraphics[width=20pc]{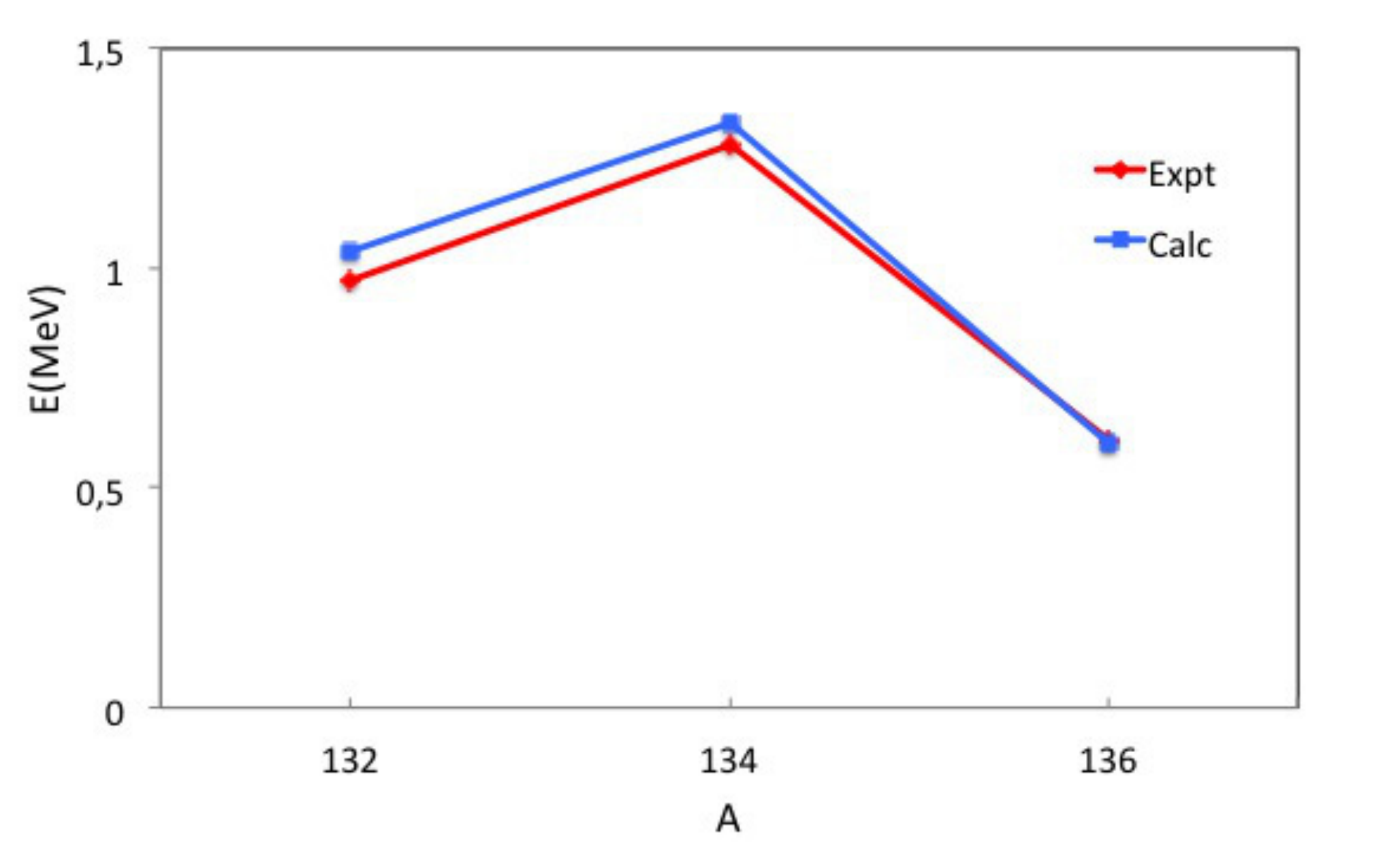}
\end{center}
\caption{\label{fig6_AG} (Color online) Calculated and experimental~\cite{NNDC} excitation energies of the yrast $2^+$ states in tellurium isotopes with $A=132$, 134, and  136.}
\end{figure}

The theoretical and experimental excitation energies of the yrast $2^+$,  $4^+$, and $6^+$ states in Sn isotopes from $A= 126$ to 136 are compared in figure~\ref{fig5_AG}, while for Te isotopes we  consider only the yrast $2^+$ states  for $A=132$, 134, and 136, as shown in figure~\ref{fig6_AG}. We see that the available experimental data are very well reproduced by our calculation, including the marked decrease in the 
energies at $N=84$. Note that for $^{136}$Sn, for which no spectroscopic information is available, we predict  a low energy spectrum quite similar to that of $^{134}$Sn.

In table~\ref{tab2} the calculated $B(E2; 2^{+}_{1} \rightarrow 0^{+}_{1})$ transition rates, obtained with an effective proton and neutron charge of 1.55 and 0.7$e$,
respectively, are compared with the experimental values.
We see that in all cases, except for $^{136}$Te, the agreement between theory and experiment is very good. More precisely, the discrepancy  ranges between  12 and 22\%, becoming larger than 50\% in $^{136}$Te. It is worth mentioning, however,  that a new  higher precision measurement  of the $B(E2)$ in $^{136}$Te is certainly needed, the present experimental value being determined from a reanalysis~\cite{Danchev11} of the experiment of  Ref. \cite{Radford02}. At the same time,  the predicted overestimation  may be seen as an indication  that our effective interaction lacks some accuracy.
Actually, we pin down the main feature of the $2^+$ state wave function, which turns out to be dominated by neutron excitations, but  find  that a 35\% remaining weight is  fragmented over  various components,  which are likely to lead to an enhancement of the B(E2).

\begin{table}[h]
\caption{\label{tab2} Experimental~\cite{Danchev11,Radford05} and calculated $B(E2; 2^{+}_{1} \rightarrow 0^{+}_{1})$ (in W.u.) for Sn and Te isotopes .} 
\begin{ruledtabular}
\vspace{0.6cm}
\centering
\begin{tabular}{lcc}
&  Expt & Calc \\
\colrule
$^{130}$Sn & 1.2(3) & 1.4  \\
$^{134}$Sn & 1.4(2) & 1.6  \\          
$^{132}$Te & 10 (1) & 7.8  \\  
$^{134}$Te & 5.6 (6) & 4.9  \\  
$^{136}$Te & 5.9 (9) & 9.9  \\  
\end{tabular}
\end{ruledtabular}
\end{table}

\begin{figure} [h]
\begin{center}
\includegraphics[width=18pc]{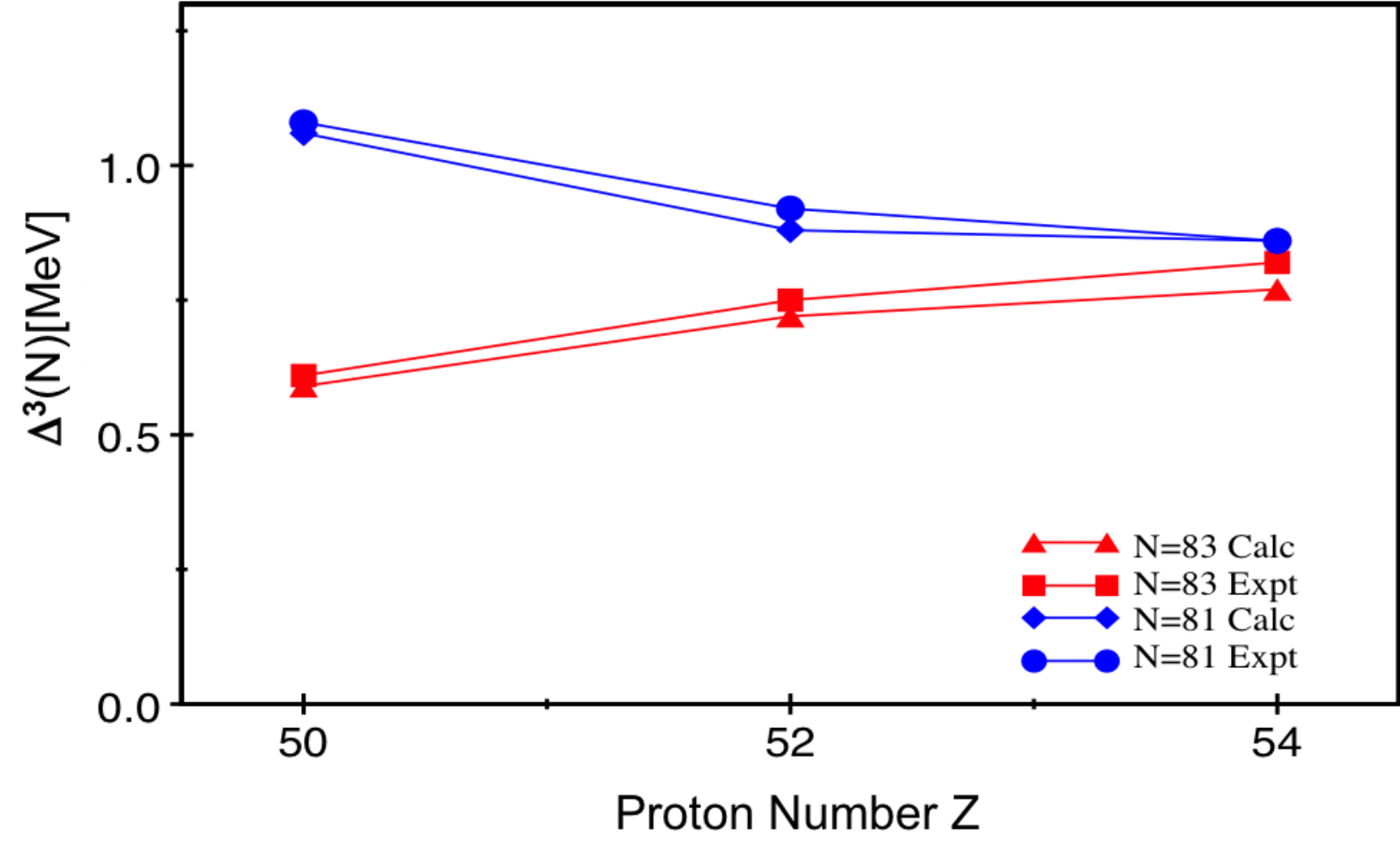}
\end{center}
\caption{\label{fig7_AG} (Color online) Calculated and experimental~\cite{Hakala12} odd-even staggering for the $N= 81$ and 83 isotones.}
\end{figure}

As mentioned above, we conclude  this section by discussing the neutron OES as defined by three-point formula

\begin{equation}
\Delta^{(3)} (N,Z)=\frac{1}{2} [B(N+1, Z) +B(N-1,Z)-2B(N,Z)].
\end{equation}
\noindent
By using our  binding  energies for $^{130,134}$Sn, $^{132-136}$Te, and $^{134-138}$Xe, we have calculated  the neutron  OES for the $N=81$  isotones  $^{131}$Sn, $^{133}$Te, $^{135}$Xe and for the $N=83$ isotones $^{133}$Sn, $^{135}$Te, $^{137}$Xe. They are compared with the experimental values in figure~\ref{fig7_AG}. 
We see that the agreement between theory and experiment is very good. In particular,
our calculations give a quantitative description of  the gap between the $N= 81$ and 83 lines at
$Z= 50$ as well as of its decrease when adding two and four protons. The drop of about 0.5 MeV in the observed OES for Sn when crossing N= 82 is a consequence of the different pairing properties  for neutron particles and holes with respect to the N= 82 closed shell. When going to Te and Xe, the $N= 81$ and 83 lines come closer to each other as a result of the proton-neutron effective interaction. The two lines would be indeed parallel should one ignore this interaction. From figure~\ref{fig7_AG}, we see that the $p−n$ interaction has an opposite effect on the $N= 81$ and $N= 83$ isotones, which is clearly related to its repulsive and attractive nature in the particle-hole and particle-particle channel, respectively. On the other hand, this effect is not very large either in $^{133,135}$Te or in
$^{135,137}$Xe, since it results essentially from the difference between the contributions of the $p−n$
interaction to the energies of the odd and neighboring even isotopes.

\section{Summary and conclusions}

In this paper, we have given a short overview of the shell-model approach with realistic effective interactions to the study of exotic nuclei.
We have reviewed the main steps and the relevant recent developments involved in the derivation of these interactions and reported
some selected results  for neutron-rich nuclei in various mass regions, as oxygen isotopes,  $N = 40$ isotones, 
and nuclei around $^{132}$Sn. Actually, we have tried to give a panoramic view of our results,  most of them already given 
 in  previous works, in order to illustrate the practical value of this approch as well as some of its limits.

The quality of our results has been highlighted in section 3 by comparison with the available experimental data. This shows that, with a few exceptions, 
observed spectroscopic properties as well as binding energies are well reproduced by the theory, proving the predictive power of realistic shell-model
calculations. We have seen, however, that, when using theoretical single-particle energies,  these calculations fail to reproduce the binding energies of the O isotopes. 
This may be traced to the lack of a three-body force, or more precisely to the contributions of this force to  the one-body term of the effective Hamiltonian, while
our results  do not seem to evidence the need of an explicit three-body term or of renormalization of the two-body interaction.

Based on our calculations,  we can conclude that nuclear structure results do not 
depend substantially  on the choice of the free $NN$ potential one starts with, when employing phase-shift equivalent low-momentum potentials. In this connection, we note
that the effective interactions for the $N=40$ isotones and $^{132}$Sn neighbors are derived from the $V_{\rm low-k}$ of the CD-Bonn potentials
with different cutoffs. We have also verified~\cite{Covello05} that moderate variations of the cutoff do not change significantly the shell-model results.
However, further study is certainly needed on this point which has been examined essentialy for interactions derived at second order in the $\hat Q$-box.

The calculation of the $\hat Q$-box, as mentioned above, is one of the most critical point in our approach. Its convergenge properties are investigated in~\cite{Coraggio12}.
Here we point out  that the quality of our results for nuclei
around $^{132}$Sn is very good when using a second-order $\hat Q$ box.  One should consider, however, that  calculations  in this region
 are limited to nuclei near shell closures and cannot exclude that effects related to third order may appear for more valence neutrons  (see comments  in~\cite{Covello10}).
To conclude,  we may say that realistic-shell model calculations  represent by now a very effective tool to study nuclear structure. This makes it challenging to try to clarify some remaining open questions.


\begin{thebibliography}{99}
\bibitem{Coraggio07} Coraggio L, Covello A, Gargano A, Itaco N, Entem D R, Kuo T T S
and Machleidt R 2007 {\it Phys. Rev.} C {\bf 75} 024311
\bibitem{Coraggio11} Coraggio L, Covello A, Gargano A, Itaco N and Kuo T T S 2011
 {\it  J. Phys. Conf. Ser.} {\bf 312}  092021
\bibitem{Coraggio14} Coraggio L, Covello A, Gargano A and Itaco N 2014 submitted to
{\it Phys. Rev.} {\bf C}
\bibitem{Coraggio13} Coraggio L, Covello A, Gargano A and Itaco N 2013 {\it Phys. Rev.}
C {\bf 88} 041304(R) and references therein
\bibitem{Machleidt01} Machleidt R 2001 {\it Phys. Rev.} C {\bf 63} 024001
\bibitem{Bogner02} Bogner S, Kuo T T S, Coraggio L, Covello A and Itaco N 2002 {\it Phys. Rev.}
C {\bf 65} 051301(R)
\bibitem{Ekstrom13}  Ekstr\"{o}m A, Baarsden G, Forss\'{e}n, Hagen G, Hjorth-Jensen M,
Jansen G R, Machleidt R, Nazarewicz W, Papenbrock T, Sarich J and Wild S M
2013 {\it Phys. Rev. Lett.} {\bf 110} 192502
\bibitem{Coraggio09} Coraggio L, Covello A, Gargano A, Itaco N and Kuo T T S 2009 {\it Prog. Part. Nucl. Phys.} {\bf 62} 135
\bibitem{Coraggio12} Coraggio L, Covello A, Gargano A, Itaco N and Kuo T T S 2012 {\it Ann. Phys. } {\bf 327} 2125
\bibitem{Suzuki89} Suzuki K and Lee S. Y. 1989 {\it Prog. Theor. Phys.} {\bf 64} 2091
\bibitem{Covello10} Covello A and Gargano A  2010 {\it J. Phys. } G {\bf 37} 064044
\bibitem{Otsuka10} Otsuka T, Suzuki T, Holt  J D, Schwenk A,  and Akaishi Y 2010  {\it Phys. Rev. Lett.} {\bf 105} 032501
\bibitem{Hagen12}  Hagen G, Hjorth-Jensen, M, Jansen G R, Machleidt R and Papenbrock T
2012 {\it Phys. Rev. Lett.} {\bf 108} 242501
\bibitem{Hergert13} Hergert H, Binder S, Calci A, Langhammer J and Roth R 2013 {\it Phys. Rev. Lett.} {\bf 110} 242501
\bibitem{NNDC} Data extracted using the NNDC On-line Data Service from
the ENSDF database, file revised as of January 8, 2014
\bibitem{AMU} Wang M, Audi G,  Wapstra A H, Kondev F G, MacCormick M, Xu X and Pfeiffer B 2012 {\it Chinese Phys.} C {\bf 36}  1603 
\bibitem{Lenzi10} Lenzi S M, Nowacki F, Poves A and Sieja K 2010 {\it Phys. Rev.} C {\bf 82} 054301
\bibitem{Rother11} Rother W {\it et al.} 2011 {\it Phys. Rev. Lett.} {\bf 106} 022502
\bibitem{Baugher12} Baugher T {\it et al.} 2012 {\it Phys. Rev.} C {\bf 86} 011305
\bibitem{Crawford13} Crawford H L {\it et al.}  2013 {\it Phys. Rev. Lett.} {\bf 110} 242701
\bibitem{Coraggio9b} Coraggio, L, Covello A, Gargano A and Itaco N 2009 {\it Phys.
Rev}  C {\bf 80} 021305(R)  and references therein
\bibitem{Covello11} Covello A, Coraggio L,  Gargano A and Itaco N 2011  {\it  J.
Phys. Conf. Ser.} {\bf 267} 01201 and references therein
\bibitem{Danchev11} Danchev M {\it et al.}  2011 {\it Phys. Rev.} C {\bf 84} 061306(R)
\bibitem{Coraggio13b} Coraggio L, Covello A,  Gargano A and Itaco N 2013 {\it Phys.
Rev.} C {\bf 87} 034309
\bibitem{Covello13} Covello A, Gargano A and Kuo T T S 2013 {\it Fifty Years of Nuclear BCS}
ed R A Broglia and V Zelevinsky (Singapore: World Scientific) pp. 169-178
\bibitem{Terasaki02} Terasaki J, Engel J, Nazarewicz W and  Stoitsov M 2002 {\it Phys. Rev.} C
{\bf 66} 054313
\bibitem{Shimizu04} Shimizu N, Otsuka T, Mizusaki T and  Honma M 2004
{\it Phys. Rev.} C {\bf 70} 054313
\bibitem{Radford02} Radford D C {\it et al.} 2002 {\it Phys. Rev. Lett.} {\bf 88} 222501
\bibitem{Radford05} Radford D C {\it et al.} 2005 {\it Nucl. Phys} A {\bf 752} 264c
\bibitem{Hakala12} Hakala J. {\it et al.} 2012 {\it Phys. Rev. Lett.} {\bf 109} 032501
\bibitem{Covello05} Covello A, Coraggio L, Gargano A and Itaco N 2005
{\it J.  Phys. Conf. Ser.} {\bf 20} 137
\end{thebibliography}
\end{document}